\documentclass[11pt, onecolumn]{article}

\usepackage{float}
\usepackage{graphics} % for pdf, bitmapped graphics files
\usepackage{epsfig} % for postscript graphics files
\usepackage[caption = false]{subfig}
\usepackage{cite}
\usepackage[cmex10]{amsmath}
\usepackage{datetime}
\usepackage{graphicx}
\usepackage{stfloats}
\usepackage{color}
\usepackage[T1]{fontenc}
\usepackage[utf8]{inputenc}
\usepackage{authblk}
\usepackage{amssymb}
\usepackage{algorithm}
\usepackage{pstricks,epsfig}
\usepackage{rotating}
\usepackage{subfig}
\usepackage{tikz}
\usepackage{enumerate}
\usepackage{tabularx}
\usepackage{verbatim}
\usepackage{pgf}
\usepackage{url}
\usepackage[margin=2.5cm]{geometry}
\usepackage{setspace}
\begin{document}

\title{Resolution- and throughput-enhanced spectroscopy using high-throughput computational slit}

\author{Farnoud~Kazemzadeh$^{1,\ast}$, and Alexander~Wong$^{1}$\\
$1$ Department of Systems Design Engineering, University of Waterloo, Waterloo, Ontario, Canada, N2L 3G1\\
$\ast$ E-mail: fkazemzadeh@uwaterloo.ca}

\date{}
\maketitle
\clearpage

\textbf{There exists a fundamental tradeoff between spectral resolution and the efficiency or throughput for all optical spectrometers. The primary factors affecting the spectral resolution and throughput of an optical spectrometer are the size of the entrance aperture and the optical power of the focusing element. Thus far collective optimization of the above mentioned has proven difficult. Here, we introduce the concept of high-throughput computational slits (HTCS), a numerical technique for improving both the effective spectral resolution and efficiency of a spectrometer. The proposed HTCS approach was experimentally validated using an optical spectrometer configured with a 200 $\mu$m entrance aperture, test, and a 50 $\mu$m entrance aperture, control, demonstrating improvements in spectral resolution of the spectrum by $\sim$50\% over the control spectral resolution and improvements in efficiency of > 2 times over the efficiency of the largest entrance aperture used in the study while producing highly accurate spectra.}

The spectral resolution and efficiency of an optical spectrometer, which affects the ability to distinguish closely spaced spectral features which are controlled by a number of different factors such as the size and shape of the entrance aperture, the optical characteristics of the dispersive element, the optical characteristics of the collimating optics the focusing optics, and the size and shape of the detector's pixels. In particular, the size of the entrance aperture (e.g., width of slit used as entrance aperture) and the optical power of the focusing element are the primary factors affecting the tradeoff between spectral resolution and efficiency or throughput of a spectrometer. A focusing element that can create a very sharp focused spot can be useful in increasing the spectral resolution of a spectrograph but, there exists fundamental limitations in manufacturing and designing of focusing elements with high optical powers (small focal ratio, f-number or large numerical aperture) therefore more efforts have been put forth in decreasing the effective size of the entrance aperture in order to increase the spectral resolution.

The spectral resolution of the spectrometer can be increased by decreasing the size of the entrance aperture. The decrease in size, width, is only necessary along the spectral dispersion axis. By decreasing the width of the entrance aperture of the spectrometer the width of the resulting focused spectral spots are also decreased therefore subtending a smaller portion of the detector array to allow closely spaced spectral features to be distinguished. However, this decrease in size of the entrance aperture also decreases the amount of light that enters the spectrometer -as less light is able to enter the spectrometer- therefore resulting in a decrease in light throughput of the resulting spectrum. The depletion of light can be detrimental to application where transient phenomena are the subject f study or systems that are photon-starved -where the amount of light available to the spectrometer is low-  yet demand extremely high spectral resolution such as Raman spectroscopy~\cite{Garrell89, Haynes05, Stiles08}, biomedical spectroscopy~\cite{Lin12, Krafft06}, and astronomy~\cite{Chu94, Appenzeller12} among many other applications. Thus, spectroscopic methods that allow for an increase in spectral resolution while maintaining and allowing for high light throughput are sought after.

Previous methods of improving both the spectral resolution and efficiency of spectrometers have focused primarily on the design of analog optical slicers situated before the dispersive element of the spectrometer. Some optical slicers use specialized prisms to slice a light beam~\cite{Bowen38}, where the performance depends on the optical properties of the prism which is wavelength dependent and can limit its use in broadband light conditions. Some optical slicers -often referred to as integral field slicers- make use of slicer mirror arrays, lenslet arrays, or fiber optical bundles, in the image space, to redirect portions of the light to their respective spectrometer, thus slicing the light beam into portions with at least one of the spatial dimensions smaller than the received light beam. However, such optical slicers can be large in size and limited in getting high spectral information from all the different beam portions~\cite{Fischer10,Gao09,Cardona10, Laurent08, Avila12}.

\begin{figure*}[!t]
	\centering
	\includegraphics[width=0.92\textwidth]{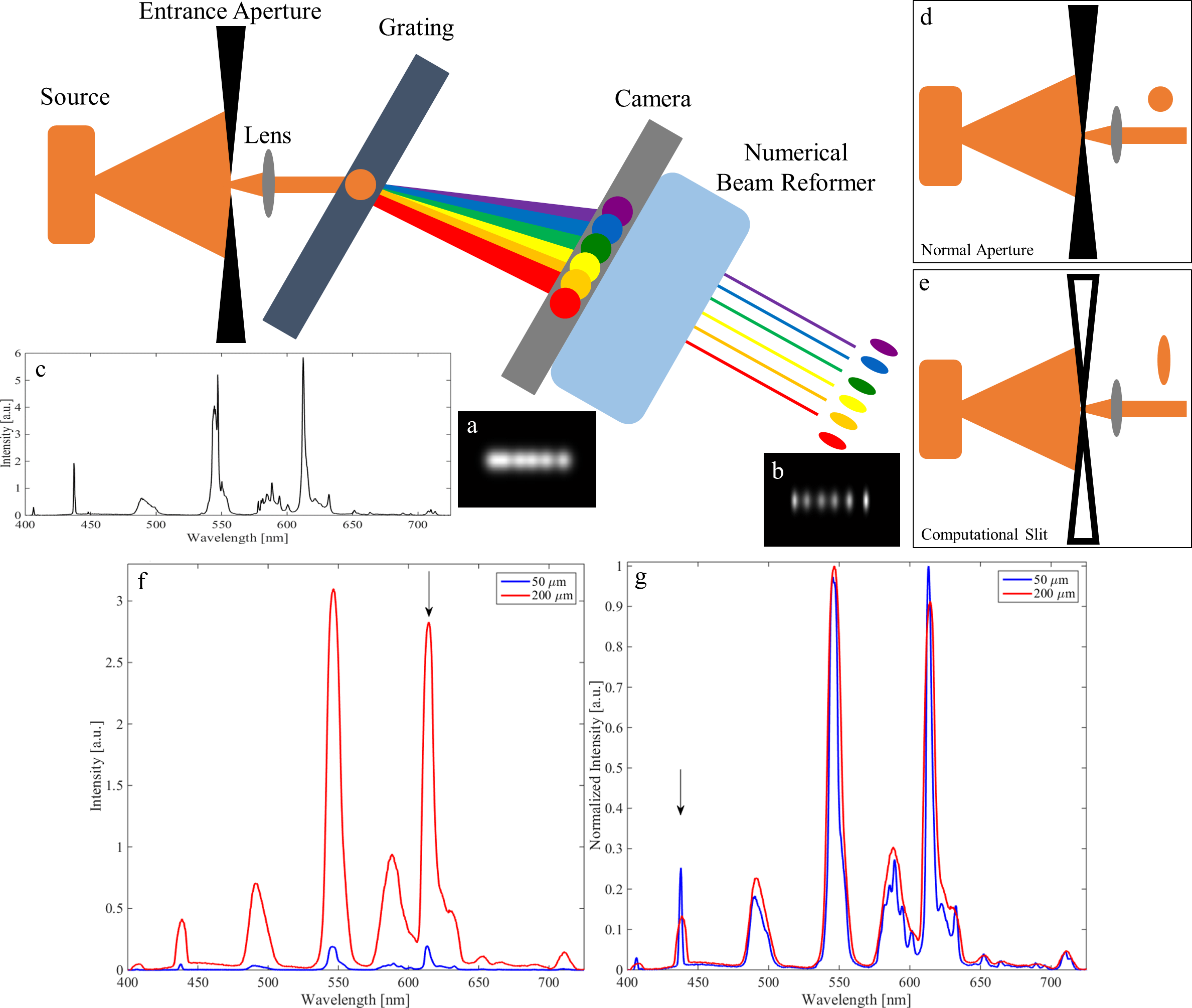}
	\caption{\textbf{Overview of high-throughput computational slit (HTCS).} The systems design of a generic spectrometer is shown here augmented with the proposed HTCS technique. Pictorial representation of the beam shape demonstrate how HTCS operates by changing the morphology of light that has entered from a much larger entrance aperture (\textbf{a)}to increase the spectral resolution of the spectrometer by imposing a small-slit function(\textbf{b)}. (\textbf{c}) The high-resolution spectrum of the light source used for this study, Mercury vapor fluorescent lamp captured using a 10 $\mu$m entrance aperture. (\textbf{d}) Shows the imaged spot morphology after the light propagates through a normal aperture. (\textbf{e}) Shows the imaged spot morphology after the light propagates through the proposed HTCS. (\textbf{f}) The spectra of the light source at one second exposure time using a 200 $\mu$m and a 50 $\mu$m entrance aperture used for light throughput comparison. It is noted at the marked wavelength of interest, 613 nm, the peak signal is > 14 times larger using the larger entrance aperture. (\textbf{g}) The normalized spectra of the source using a 200 $\mu$m and a 50 $\mu$ m entrance aperture used for spectral resolution comparison. It is noted at at the marked wavelength of interest, 438 nm, the spectral resolution is > 4 times higher using the smaller entrance aperture.}
	\label{fig1}
\end{figure*}

In~\cite{Meade13}, Meade et al. introduced a pupil-based optical slicer, comprising of a beam reformatter, a beam compressor and a beam expander to improve spectral resolution while allowing for high throughput via a large entrance aperture by negating the use of a light limiting slit. The beam reformatter receives a full aperture of collimated light beam and splits it into two or more beam portions in the dimension of dispersion. These beam portions, each containing similar spectral information, are then reformed into a vertical stack which has a narrower width in the direction of dispersion. An asymmetric beam expander then changes the width of the stack to match the original width of the original full aperture. While such optical slicing methods do improve both spectral resolution and throughput, they come at the expense of the introduction of additional analog optical elements that increase not only the complexity of the spectrometer, but also increases the cost of the spectrometer as well~\cite{Meade14}. These optical components (e.g., lenses, reflective surfaces, etc.) along with the associated mounting apparatus will introduce aberrations and other performance issues (i.e. alignment) to the device. Therefore, a method that allows for an increase in both spectral resolution and throughput without the introduction of additional analog optical components and associated mounting apparatus, such as that in an optical slicer, to the spectrometer is highly desired.

\begin{figure*}[!t]
	\centering
	\includegraphics[width=1\textwidth]{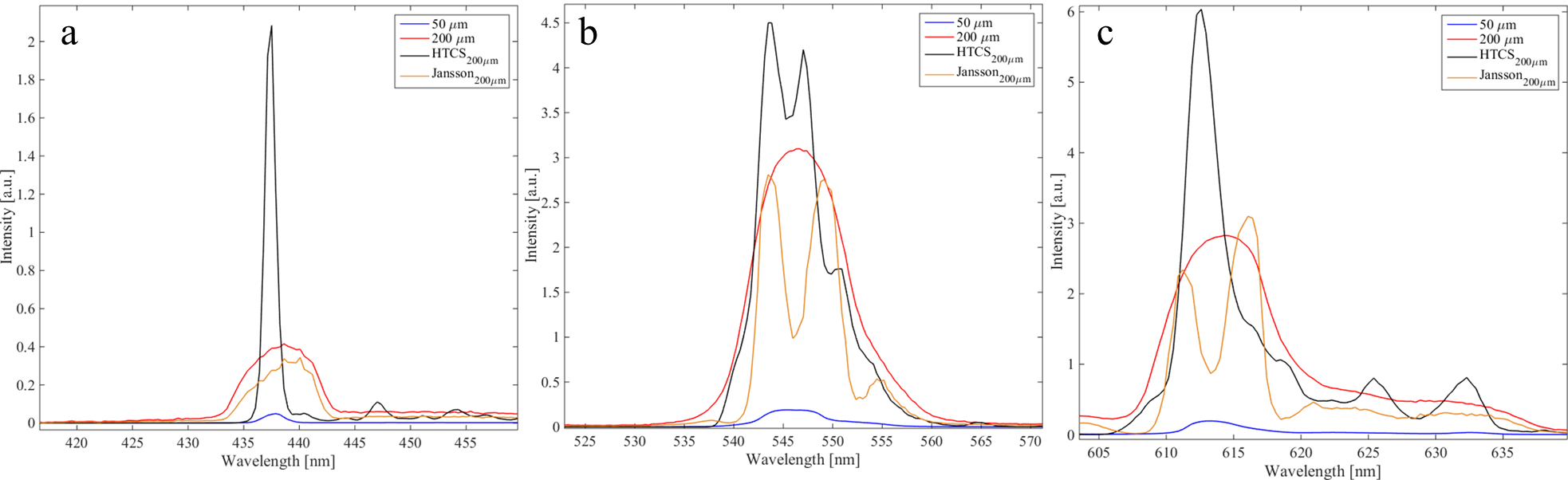}
	\caption{\textbf{The spectra of the light source at one second exposure time using a 200 $\mu$m and a 50 $\mu$m entrance aperture, and using HTCS and Jansson deconvolution on the 200 $\mu$m entrance aperture spectrum demonstrating the improvements in spectral characteristics resulting from HTCS.} (\textbf{a}) The improvement in spectral resolution as well as the signal-to-noise ratio of the spectra are demonstrated. The spectral resolution is calculated to be 4.22 nm, 1.35 nm, 3.5 nm, and 0.69 nm for the 200 $\mu$m aperture, 50 $\mu$m aperture, Jansson deconvolution, and HTCS, respectively. The SNR of the spectra are also calculated be 27.66 dB, 35.88 dB, 29.01 dB, and 40.99 dB for the 200 $\mu$m aperture, 50 $\mu$m aperture, Jansson deconvolution, and HTCS, respectively. (\textbf{b}) The doublet feature of the spectrum which cannot be observed in the spectra measured using the 200 $\mu$m and the 50 $\mu$m entrance apertures; however, it is accurately reproduced by the HTCS and inaccurately produced by the Jansson deconvolution method. (\textbf{c}) The spectral line is inaccurately reproduced as a doublet by the Jansson deconvolution method.}
	\label{fig2}
\end{figure*}

In this study, we introduce the concept of high-throughput computational slit (HTCS) for improving both the effective spectral resolution and efficiency of the resulting spectrum. The HTCS technique is a numerical technique that is performed after the spectral information is captured by the detector.  An overview of the proposed concept of HTCS is shown in Fig.~\ref{fig1}. In HTCS light enters an entrance aperture (pinhole), similar to any other spectrometer, however the size of the aperture need not to be limited by the desired spectral resolution, thus, increasing the amount of light that enters the system. The light diverging out of the entrance aperture encounters a collimating element to produce a collimated light beam. The collimated light beam then continues on and is then projected onto a dispersive element, which splits the input light beam into its spectral components in the form of multiple collimated light beams traveling at different angles depending on the wavelength of light. The spectral beams then arrive at a camera which focuses the light using a focusing element and records the information on a detector. The detector recordings are then used to form digital representations of the beams that entered the detector, and the digital beams are then reformed numerically such that they appear as to have passed through a slit with the desired morphology, imposed by the spectral resolution requirement, in the first place.  By passing the digital beams numerically through a high-throughput computational slit that is narrower than the physical entrance aperture of the spectrometer, improved spectral resolution can be achieved while reaping the light throughput benefits of a much larger entrance aperture.

Figure~\ref{fig1}a, shows the image of the dispersed light that is formed on the detector, from which the spectrum is extracted, note that since the size of the input aperture is not limited in the dispersion direction there is an overlap of the spectral features which tends to decrease the spectral resolution. In other words the uniqueness of wavelengths is not discernible. But given the HTCS method, a higher-resolution spectrum can be produced, resulting in highly separable spectral features, Figure~\ref{fig1}b.

In this study a Mercury vapor fluorescent lamp was used as the source, the high-resolution spectrum of which is shown in Figure~\ref{fig1}c. The proposed HTCS operates analogous to changing the morphology the entrance aperture of the system from a circular aperture, Figure~\ref{fig1}d, to an elliptical aperture, Figure~\ref{fig1}e, in this case. The spectrum of the lamp was measured using a spectrometer with a 200 $\mu$m and a 50 $\mu$m entrance aperture and the resulting exposure time independent spectra are shown in Figure~\ref{fig1}f. It is clear that with larger entrance aperture more light is allowed to enter the spectrometer hence producing a spectrum with much higher signal. Conversely, a small entrance aperture would result in a spectrum of improved spectral resolution, Figure~\ref{fig1}g (normalized spectra), as more spectral features can be observed. Quantitatively, using the spectral feature at $\sim$ 613 nm in Figure~\ref{fig1}f, it is observed that the signal of the spectrum from the larger aperture is > 14 times higher than the signal from the smaller aperture. Additionally, according to the spectral feature at $\sim$ 438 nm in Figure~\ref{fig1}g, the spectral resolution is > 4 times higher for the spectrum of the smaller aperture, 7 nm for 200 $\mu$m and 1.7 nm for 50 $\mu$m entrance apertures. This is the tradeoff that has plagued the field since its inception.

In the HTCS method, let $I^{\alpha}$ denote the predicted beam that would have been projected on the detector if it had passed through the desired computational slit, and let $I^{\beta}$ denote the beam projected on the detector having passed through the physical slit.  We treat the problem of numerical beam reforming as a probabilistic state prediction problem, where the goal being to predict the most probable $\hat{I^{\alpha}}$ given $I^{\beta}$, computational slit function $S$ that models the shape of the beam exiting a computational slit with the desired morphology, and optical transfer function $O$.  Modeling $I^{\alpha}$ and $I^{\beta}$ as stochastic processes, the probabilistic state prediction problem can be formulated as

\vspace{-3mm}
\begin{equation}
	\hat{I^{\alpha}} = {\rm argmax}_{I^{\alpha}}~p\left(I^{\alpha} | I^{\beta}, S, O \right),
\label{MAP}
\end{equation}

\noindent where $p\left( I^{\alpha} | I^{\beta}, S, O \right)$ is the conditional probability of $I^{\alpha}$ given $I^{\beta}$, $S$, and $O$.

Accounting for quantum photon emission statistics, the morphology of the computational slit, and nonstationary of $I^{\alpha}$ and $I^{\beta}$, we introduce the following $p(I^{\alpha}|I^{\beta}, S, O)$:

\begin{footnotesize}
\begin{equation}
	p\left(I^{\alpha} | I^{\beta}, S, O\right) = \prod_{x \in X} \frac{\left(\mathfrak{F^{-1}}\left\{\frac{O}{S}\mathfrak{F}\left\{I^{\alpha}_x\right\}\right\}\right)^{{I^{\beta}_x}}e^{-\left(\mathfrak{F^{-1}}\left\{\frac{O}{S}\mathfrak{F}\left\{I^{\alpha}_x\right\}\right\}+\frac{\left(I^{\alpha}_x-{E}[I^{\alpha}_x]\right)^2}{2 E\left[(I^{\alpha}_x-E[I^{\alpha}_x])^2\right]}\right)}}{{I^{\beta}_x}!}
\label{likelihood}
\end{equation}
\end{footnotesize}

\noindent where $X$ denotes a set of sensor locations in the detector, $x \in X$ denotes a specific sensor location in the detector, and ${E}$ denotes nonstationary expectation.  In this study, an iterative EM solver is used to solve Eq.~\ref{MAP}, and is performed until convergence.

For this study, a computational slit function $S$ corresponding to a 5$\mu$m slit was used.  By utilizing a computational slit with a width narrower than the physical entrance aperture, the idea is that the spectral resolution of the resulting spectrum would be enhanced while maintaining the high light-throughput enabled by the large entrance aperture.

The proposed HTCS technique, implemented as described above, was examined on the spectrum of the fluorescent lamp using the 200 $\mu$m entrance aperture the results of which are shown in Figure~\ref{fig2} which demonstrates that not only the spectrum resulting from the HTCS technique has a higher power than the original 200 $\mu$m spectrum but it has higher spectral resolution than the 50 $\mu$m spectrum. The Jansson deconvolution technique~\cite{Jansson84} was also evaluated as a comparison to the proposed HTCS technique, with the parameters of the Jansson technique empirically tuned to minimize noise amplification while maintaining improvements in spectral resolution.

The exposure time independent spectra are shown in Figure~\ref{fig2}a which demonstrates that the signal of the HTCS spectrum is many times higher than that of the 200 $\mu$m and 50 $\mu$m spectra. The signal-to-noise ratio of the spectra are calculated to be 27.66 dB, 35.88 dB, 29.01 dB, and 40.99 dB for the 200 $\mu$m aperture, 50 $\mu$m aperture, Jansson deconvolution, and HTCS, respectively. The signal-to-noise ratio of the HTCS spectrum is 5.11 dB, 11.98 dB, and 13.33 dB higher than that of the 50 $\mu$m, Jansson, and the 200 $\mu$m spectra.  As such, it can be observed that HTCS can provide significant improvements in the SNR of the resulting spectra.

The spectral resolution of the spectra are measured by evaluating the standard deviation or the full-width-half-maximum of the Gaussian profile fit to the emission line in Figure~\ref{fig2}a. The spectral resolutions are calculated to be 4.22 nm, 1.35 nm, 3.5 nm, and 0.69 nm for the 200 $\mu$m aperture, 50 $\mu$m aperture, Jansson deconvolution, and HTCS, respectively. As such, the spectral resolution of the HTCS spectrum is $\sim$2 times, $\sim$5 times, and $\sim$7 times higher than the spectral resolution of the 50 $\mu$m, Jansson, and the 200 $\mu$m spectra, respectively. This resolution enhancement can be observed in the portion of the spectrum shown in Figure~\ref{fig2}b where a doublet feature at $\sim$ 545 nm is examined closely. This feature is not resolved in the 200 $\mu$m spectrum or the 50 $\mu$m but it is clearly resolved using the HTCS and the Jansson method. However, the Jansson method is thrown off by the spectral feature at 551 nm which tends to mix with the second peak of the doublet and redshift this peak, rendering the spectrum unreliable. Furthermore, upon closer examination of the feature at 613 nm, Figure~\ref{fig2}c, it can be observed that the Jansson deconvolution technique falsely interprets this single emission line as a doublet, thus further decreasing the reliability and fidelity of the spectrum.

Lastly, in order to demonstrate that the HTCS technique does not result in adverse effects on the fidelity of the measured spectrum, the correlation coefficient between all of the measured spectra and the ground truth spectrum presented in Figure~\ref{fig1}c was evaluated. The correlation coefficient which determines the statistical relationship, likeness, between two entities is evaluated to be 94.13\%, 87.69\%, 91.49\%, and 97.83\% for the spectra measured using the 50 $\mu$m entrance aperture, 200 $\mu$m entrance aperture, Jansson deconvolution, and HTCS compared to the ground truth spectrum presented in Figure~\ref{fig1}c. This demonstrates that not only does HTCS improve the throughput and spectral resolution of a spectrometer, it preserves the underlying features of a spectrum and is capable of reproducing highly accurate spectra, thus enabling reliable spectrometry for photon-starved or of transient phenomena.

The proposed computational technique has been demonstrated to achieve a $\sim$7-fold increase in the spectral resolution which is analogous to improving the performance of an R=77 spectrometer to that of an R=471 spectrometer. While increasing the resolution of the spectrograph the efficiency of the device is also improved as seen by the improvement of the signal-to-noise ratio by $\sim$13 dB.  A limitation of the proposed HTCS technique is that it relies on an accurate model of the optical transfer function to perform well, and as such deviations and errors in the modeled optical transfer function relative to the true optical transfer function will result in inaccuracies in the resulting spectrum produced by HTCS.  Furthermore, it is not comprehensively determined that this improvement is without limitations as the connection and effect between the entrance aperture function, optical transfer function, signal-to-noise ratio, and the linearity of the spectrum is not well studied and is beyond the scope of the current study.  For example, while the proposed HTCS technique performs well when tasked with extracting spectral features, one potential limitation that may arise is that, like existing deconvolution techniques, it may not be able to handle continua well as it may force spectral features that may not exist.

\section*{Acknowledgments}
This work was supported by the Natural Science and Engineering Research Counsel of Canada (127791), and Canada Research Chairs program.The authors would like to thank their industry partner, Lumalytics Inc.

\section*{Author Contribution}
A.W. and F.K. have equally contributed to this work.

\end{document}